\documentclass{article}

\usepackage{arxiv}

\usepackage[T1]{fontenc}    
\usepackage{hyperref}       
\usepackage{url}            
\usepackage{booktabs}       
\usepackage{amsfonts}       
\usepackage{nicefrac}       
\usepackage{microtype}      
\usepackage{lipsum}
\usepackage{graphicx}
\usepackage{multirow}
\usepackage{color}
\usepackage{enumerate}
\usepackage{lineno}
\graphicspath{ {./images/} }

\hypersetup{
colorlinks=true,
linkcolor=black,
citecolor=black
}

\title{A technical solution for the rule of law, peace, security, and evolvability of global cyberspace - solve the three genetic defects of IP network}

\author {
Hui Li$^{a*}$, Kedan Li$^{b}$, Jiaqing Lv$^{a}$, Yuanshao Liang$^{a}$, Feng Han$^{c}$, Shuo-Yen Robert Li$^{d}$ \\
$^{a}$ Shenzhen Graduate School, Peking University, China\\
$^{b}$ Department of Computer Science, University of Illinois at Urbana Champaign, USA\\
$^{c}$ Institute for Advanced Study, Tsinghua University, China\\
$^{d}$  National Key Laboratory of Wireless Communication, University of Electronic Science and Technology of China, China\\
*Corresponding author: lih64@pkusz.edu.cn
}

\begin{document}
\maketitle

\abstract{Since its inception in the 1960s, the internet has profoundly transformed human life. However, its original design now struggles to meet the evolving demands of modern society. Three primary defects have emerged: First, the concentration of power among a few dominant entities has intensified international conflicts and widened the technological divide. Second, the Internet Protocol (IP)-based system lacks inherent security, leading to frequent global cybersecurity incidents. Third, the rigidity of the IP protocol has hindered the sustainable development of cyberspace, as it resists necessary adaptations and innovations. Addressing these issues is crucial for the future resilience and security of the global digital landscape.

To address these challenges, we propose the Co-governed Multi-Identifier Network (CoG-MIN briefly as MIN), a novel network architecture that leverages blockchain technology to ensure equal participation of countries worldwide in cyberspace governance and the rule of law. As a next-generation network system, CoG-MIN integrates mechanisms such as user authentication, data signatures, and encryption to significantly enhance network security. In testing environments, CoG-MIN has consistently withstood extensive attacks during various international cybersecurity competitions. Additionally, CoG-MIN supports the evolution and interoperability of different identifier systems, remains IP-compatible, and facilitates a gradual transition away from IP, providing an adaptable ecosystem for diverse network architectures. This adaptability fosters the development and evolution of diverse network architectures within CoG-MIN, making it a natural progression for the internet's future development.

We further introduce a trilogy of cyberspace security theorems. The first theorem demonstrates that deterministic security within the current IP framework is unsolvable. The second posits a secure solution for future cyberspace architectures, offering a quantifiable exponential increase in security. The third conjecture integrates blockchain with the future network, management, legal frameworks, and insurance mechanisms to potentially guarantee the long-term rule of law, peace, and security in global cyberspace. This approach may effectively end relentless cyber warfare between national cyber forces, offering robust solutions for future cyberspace governance.

This proposal aims to guide the global internet into a new era. In China, MIN is about to enter the commercial stage and has gradually become a standard. We look forward to working with global stakeholders to build a cyberspace United Nations characterized by enduring rule of law, peace, and security.}




\maketitle

\section{Introduction}\label{sec1}

\subsection{Three major genetic defects of the IP network system}
In the evolution of communication technology, the rise of the IP network is undoubtedly a part that deserves in-depth discussion. Initially, this network was founded by academics passionate about knowledge sharing, including professors and researchers from universities, with Professor Einar Stefferud standing as a prominent figure in this field \cite{rose1993challenges}. They envisioned an open system designed to enable the free exchange of ideas and information, serving as a conduit for people to communicate and interact. At that time, the internet shone like a gem, symbolizing the brilliance of academic freedom and the spirit of collaboration.

However, this ideal state was short-lived. In 1998, 150 delegates from across the globe gathered in Reston, Virginia, to deliberate on a landmark proposal: the creation of a global governing body to oversee the assignment of domain names and internet addresses. This event is often seen as the constitutional moment of the internet\cite{mueller2009ruling}. The formation of the Internet Corporation for Assigned Names and Numbers (ICANN) marked a shift toward centralized governance of the internet. This shift introduced a powerful current that challenged the original ideals of freedom and democracy underpinning the internet, gradually consolidating control within a few organizations and states. Many interpret this change as a transition from the democratic principles of the internet to a form of feudal autocracy, deepening the problem of unilateral monopolies in internet governance\cite{zittrain2009future}. Throughout this process, developed countries and their tech giants have dominated resource allocation, leaving developing countries with limited influence on global cyber governance. This imbalance not only diminishes the potential for global cooperation, but also intensifies the technological divide\cite{denardis2014global}.

Meanwhile, the inherent lack of security in IP systems has exacerbated cybersecurity issues. Vinton Cerf, one of the founders of IP, acknowledged that because early networks primarily served insiders, the presence of malicious users was nearly overlooked, resulting in minimal attention to security\cite{cerf}. The IP architecture's primary goal is to route and deliver network packets, without provisions for user identity verification or packet traceability. As the Internet became public, the increase in malicious users intensified these security concerns. Within the IP system, there is no reliable way to verify the integrity and authenticity of packets, allowing malicious actors to forge source addresses and packets easily, stealing or altering data through man-in-the-middle attacks. Furthermore, distributed denial of service(DDoS) and Domain Name System(DNS) spoofing attacks threaten the stability of network services and user access security. Inevitable vulnerabilities in IP systems have led to frequent cyberattacks, with the 2017 WannaCry ransomware attack being a notable example. This attack quickly spread to over 150 countries, resulting in billions of dollars in economic losses and posing severe risks to user data and privacy\cite{zheng2017wannacry}.

IP networks have significantly constrained the diversification and technological innovation of network architectures. From the 1970s to the 1990s, various network architectures emerged globally, such as Asynchronous Transfer Mode (ATM)\cite{kim1995atm}, X.25\cite{barrett1991lan}, and Frame Relay\cite{grossman1991overview}. However, as IP networks gained popularity and dominance, these architectures were gradually phased out because of their inability to compete with IP networks. With the explosive growth in demand for video content, traditional transmission mechanisms began to show limitations, prompting academia and industry to reevaluate current network architectures and explore more adaptive solutions, such as Content-Centric Networking (ICN)\cite{de2013information}. However, the rigid IP protocol architecture has become a barrier to these innovations. Furthermore, the IP protocol itself shows evolutionary limitations. Despite more than 20 years of development, IPv6 has yet to broadly replace IPv4, reflecting the severe enlargement of IP networks. This intrusion restricts the flexibility and innovative capacity of the network ecosystem, ultimately hindering the growth of network diversity.

\subsection{A solution Co-governed Multi-Identifier Network}
In the digital age, the internet was once celebrated as a bastion of freedom, epitomizing openness and democracy. However, the rise of network centralization and monopolistic structures has placed these ideals under significant strain. To counter this challenge, we propose the Co-governed Multi-Identifier Network (CoG-MIN), an innovative network architecture designed to establish a new framework for the rule of law, peace, security, and ecosystem evolution in cyberspace. CoG-MIN transcends mere technological advancement; it embodies a transformative vision for the internet's future.

The primary goal of MIN is to dismantle existing monopolistic structures, redefine internet operations, and restore the foundational values of freedom and democracy. Central to this architecture is the Multi-Identifier System (MIS)\cite{MIS}, which moves away from centralized management of IP addresses and domain names. Operating on a hierarchical consortium blockchain, MIS enables countries to collectively manage top-level identifiers through a voting mechanism. Within each nation, a scalable hierarchical blockchain allows for autonomous network identity management. 

At the heart of MIS is its unique consortium blockchain consensus algorithm, PPoV\cite{wang2021data}, which addresses the CAP theorem's trilemma\cite{wang2024gbt} for distributed systems, achieving optimal complexity and theoretical efficiency. Testing across 200 server nodes in 26 countries on five continents has demonstrated that MIS's consensus algorithm enables powerful real-world performance. MIS represents a historic shift in global cyberspace management, ensuring democratic governance across the internet.

Endogenous security is another critical feature of MIN. Unlike traditional data transmission systems, MIN establishes a secure and traceable network environment. Within the Multi-Identifier System (MIS), each user is assigned a cryptographic identity linked to their real information and biometrics. Through a multilayered security mechanism that encompasses user authentication, data encryption, and digital signatures, MIN ensures the integrity and confidentiality of data in transit\cite{lin2022identity}. When data is transmitted, each packet is signed with the user's private key, ensuring that all transfers are encrypted and authenticated. 

MIN functions as an invisible security shield, protecting personal information from theft and preventing unauthorized data tampering. In addition, all activity logs are recorded on the blockchain, providing robust traceability. Given the frequency of cybercrime, this traceability is especially vital, enabling administrators to quickly identify malicious actors and significantly reduce security maintenance costs. MIN has been rigorously tested as a target environment in numerous international security competitions, consistently withstanding attacks from top global teams with its bounties remaining untouched. This proven resilience affirms MIN's security capabilities, ensuring a safe digital journey for users in cyberspace.

Furthermore, MIN supports the compatibility and evolution of various network identifiers, offering a comprehensive ecosystem for the future development of network architecture. Designed to dismantle network architecture monopolies and rigidity, MIN is the first architecture capable of IP compatibility and eventual IP phasing-out. MIN's network packets use TLV (Type-Length-Value) encoding, enabling continual updates and evolution. Supporting both Push and Pull semantics, MIN is compatible with a range of ICN architectures.

Within MIN, a scalable multi-identifier management scheme and a universal addressing scheme based on scalable multi-identifiers\cite{wang2021scalable} allow for the dynamic evolution of identifiers and communication models at the network layer, as well as future identifiers, routing schemes, and packet formats\cite{enl}. MIN's role in packet-switched networks is similar to the No.7 signaling system in telecommunications, making MIN the integrative platform for the advancement of global network architecture\cite{li2021co}. Under MIN's orchestration, different identifiers can co-exist harmoniously, like instruments in a symphony, coordinated and distinct. By empowering network devices with adaptive evolution capabilities, MIN significantly reduces the economic and environmental costs associated with network architecture upgrades.

\subsection{Ensuring the rule of law, peace, security, and evolvability of global cyberspace with MIN}
Since its introduction in 2019, the MIN architecture has garnered significant attention from both academia and industry. In its inaugural year, the MIN prototype received the Leading Scientific and Technological Achievements Award at the World Internet Conference in Wuzhen the first prototype ever to be honored at this prestigious event. The following year, Tsinghua University Press published the first monograph on MIN in Chinese. In 2021, MIN's inaugural English monograph, Co-governed Sovereignty Network: Legal Basis and Its Prototype \& Applications with MIN Architecture, was published by Springer Nature.

The momentum surrounding MIN continued to grow in subsequent years, demonstrating its relevance and adaptability across a variety of platforms and events. In October 2022, MIN was awarded the Diamond Award as the Global Champion at the British Invention Show. In July 2023, the architecture was prominently featured at the World Artificial Intelligence Conference. This momentum carried into November 2023, with a keynote presentation on MIN architecture delivered at the IEEE International Organization Side Event during the World Internet Conference in Wuzhen. In April 2024, at the invitation of UNCSTD Chairman Peter Major, MIN was presented to a global audience at the 27th UN CSTD Annual Meeting in Geneva \cite{un}. Two of the largest telecom operators in China have adopted MIN as a technology solution. Several groups have adopted MIN as their communication standard. MIN is being used more and more widely.

MIN's academic contributions have also expanded through detailed technical analyses. Two monographs currently under review by Springer Nature, IEEE, and Wiley delve into MIN's technical solutions and its novel approach to addressing the CAP trilemma in blockchain consortiums. These works are poised to advance the academic understanding of MIN's groundbreaking contributions, further positioning it as a transformative force in network architecture.

Drawing from the current challenges in network security and the pressing need for innovation, we propose a trilogy of security theorems to guide the evolution of cyberspace. Theorem 1 posits that it is technically impossible to achieve complete security within IP-based cyberspace, highlighting the limitations of the current framework. Theorem 2, using IP-based systems as a reference, suggests that future cyberspace architectures can exponentially improve security through technical advancements. Conjecture 3 further envisions deterministic security solutions, offering a clear direction for the advancement of network security.

Overcoming these challenges requires a unified and innovative approach, leveraging the fragmented insights of past research into a cohesive framework. In response, we propose CoG-MIN as a unifying architecture, likened to threading scattered pearls into a harmonious “necklace.” By synthesizing disparate contributions into a single cohesive framework, CoG-MIN offers a robust vision for the future of network architecture.

Guided by the foundational principles of MIN, this architecture emphasizes the rule of law, peace, security, and ecological evolution in cyberspace. Importantly, it facilitates a smooth transition from IP to non-IP networks, driven organically by market forces and user demand rather than enforced adoption. While IP is likely to remain dominant in regions such as the United States, MIN provides a viable alternative for nations seeking to assert cyberspace sovereignty \cite{li2021co}. By uniting these efforts under the umbrella of MIN, this architecture addresses the limitations of existing systems while fostering global collaboration and innovation in cyberspace governance. Through this integrated approach, MIN sets the stage for a more secure, equitable, and sustainable digital future.

\section{Three major defects in cyberspace}\label{sec2}

In today's highly interconnected world, cyberspace has become a vital domain for national development and global governance. Despite its transformative potential, this domain faces profound challenges that hinder its secure and sustainable evolution. This section delves into three critical defects of the IP network architecture and examines their far-reaching negative impacts on individuals, organizations, and nations.

\subsection{Cyberspace is monopolized unilaterally}

A unilateral monopoly dominates cyberspace, largely controlled by a small number of developed nations and major technology corporations. This concentration of power obstructs the participation of most countries in global network governance, creating a substantial void in international oversight. Such monopolistic practices deepen the technological divide between nations and intensify conflicts within the digital realm.

The concept of unilateral monopoly in cyberspace encompasses the control exercised over network infrastructure and resource allocation by a select group of developed nations and technology conglomerates. The allocation of IP addresses, for instance, highlights this disparity. While five nonprofit Regional Internet Registries (RIRs) globally oversee these allocations, developed nations, particularly the United States, disproportionately dominate these resources. This imbalance underscores deficiencies in international governance, leaving developing nations grappling with significant obstacles to building and modernizing their internet infrastructure.

Domain name management further illustrates this disparity. It is centralized within the Internet Corporation for Assigned Names and Numbers (ICANN), a nonprofit entity headquartered in the United States. Although ICANN espouses a global outlook, its decision-making has historically been influenced by the U.S. government, limiting the authority of other nations. This centralization restricts equitable participation in internet governance and perpetuates technological inequities.

The structure of the Domain Name System (DNS) root servers amplifies this dependency. Of the 13 groups of root servers, the majority are located within the United States, with only a few outside its borders. This arrangement constrains the autonomy of other nations in addressing cyber threats and technical challenges, fostering reliance on U.S.-based support. This dependence not only undermines sovereignty but also limits the ability of nations to engage effectively in international cyberspace governance.

In his groundbreaking work, Harvard Professor Jonathan Zittrain explores the profound implications of power concentration in cyberspace\cite{zittrain2009future}. He highlights how these unilateral monopolies can prioritize the interests of a privileged few over the broader public needs, creating a critical gap in global internet governance. As a result, nations are limited to implementing regional regulations, leading to a fractured digital landscape marked by significant variations in cybersecurity protocols and data protection strategies across different jurisdictions. This systemic lack of international coordination ultimately weakens cyberspace's resilience, creating fertile ground for malicious actors to strategically execute cyberattacks and pursue criminal endeavors.

The “UN Convention Against Cybercrime,” adopted by the UN Ad Hoc Committee on Cybercrime on August 8, 2024, represents a significant milestone in global cyber governance by establishing an international legal framework to combat cybercrime and regulate data access. However, its focus on cybercrime leaves broader issues of state participation in international cyberspace governance largely unaddressed. Joseph Nye, former U.S. Assistant Secretary of Defense for International Security Affairs, underscores the risks posed by the absence of effective mechanisms for international cyberspace governance. This void may compel nations to act unilaterally, heightening the risk of cyber-conflicts \cite{nye2010cyber}. For instance, the establishment of the U.S. Cyber Command in 2010 exemplifies how the lack of global governance has positioned cyberspace as a potential battlefield \cite{wiki:United_States_Cyber_Command}.

The geopolitical interests of major powers have further shaped internet governance, turning cyberspace into an arena of political contention. Lucas Kello, Associate Professor at the University of Oxford, highlights the vigorous pursuit of digital weaponry, including malware and hacking tools, by nations, exacerbating distrust and tension \cite{kello2013meaning}. The anonymity inherent in cyberattacks, facilitated by IP networks, complicates attribution, leading to misinterpretations and escalating geopolitical conflicts \cite{kello2017virtual}.

Another critical consequence of cyberspace's unilateral monopolies is the widening technological divide. The disparities in technological and digital infrastructure between developed and developing nations have deepened the economic and social inequalities. Laura DeNardis, a Yale University researcher, argues that the digital divide extends beyond technical limitations, carrying profound social and political implications that marginalize certain nations from fully participating in the digital age \cite{denardis2014global}. Additionally, the market dominance of multinational technology corporations undermines both digital diversity and democratic values. Their monopolistic practices stifle the growth of technological startups in developing nations, limiting innovation and, in some cases, exerting undue political influence over other countries.

\subsection{The IP system lacks genetic security}

The origins of today's Internet protocols can be traced back to the 1970s, originally designed to facilitate communication between computers. The primary objective is to ensure reliable data transfer between various networks. The "security gene" of a network implies that all network packets are traceable and encrypted. However, IP networks are devoid of such security features. This oversight in design has led to inherent vulnerabilities in network security, triggering continuous global cybersecurity incidents, resulting in substantial economic losses, and posing significant challenges to the protection of user privacy.

In its early days, the internet primarily served a closed community, with minimal consideration for security threats. Vint Cerf, one of the architects of the Internet Protocol (IP), acknowledged that security was not a primary focus during its design \cite{cerf}. As the Internet expanded to include the general public, it exposed critical vulnerabilities in its architecture and introduced a host of security challenges.

Data packets traverse the network without guarantees of integrity or authenticity, making them susceptible to manipulation. Malicious actors can spoof packet source addresses, enabling man-in-the-middle attacks to intercept or modify data in transit. They can also launch DDoS attacks and manipulate DNS records, destabilizing network services and threatening user security. Many existing protocols and systems were not originally built to anticipate or mitigate such threats, making them particularly vulnerable to exploitation.

Although security protocols such as IPsec\cite{dhall2012implementation} and SSL/TLS have been developed in recent years to enhance IP network protection, these measures are not infallible. Professor Ross Anderson from the University of Cambridge has noted that these intricate security protocols require specialized knowledge and specific software support. Mismanagement or user oversight can lead to the collapse of security defenses\cite{anderson2020security}. Furthermore, the inherent narrow-waist architecture of the IP system limits the efficacy of technologies implemented above the network layer to fully address security concerns. The swift evolution of network layer technologies continues to introduce new vulnerabilities, thus complicating the security landscape of IP networks. Ultimately, the IP architecture has become the weakest link in network security.

The inherent security vulnerabilities in IP networks have led to a rising frequency of global network breaches, causing significant harm not only to individuals but also to national security worldwide. Between 2018 and 2021, Facebook experienced 11 major data breaches, exposing the personal information of more than 2.5 billion people \cite{hua2024security}. These incidents increased the risks of identity theft and fraud while deepening concerns about the protection of personal privacy. In 2023, General Electric fell victim to a cyberattack that compromised numerous U.S. military secrets \cite{ge}. This breach not only tarnished GE's reputation but also posed a serious threat to U.S. national security. Despite its access to some of the most advanced cybersecurity technologies, the United States continues to grapple with persistent and evolving cybersecurity challenges.

Renowned cybersecurity expert Kevin Mitnick has highlighted that the root of cyber threats lies in the actions of malicious actors, not in the technology itself \cite{mitnick2003art}. Early IP networks lacked robust mechanisms for verifying user identities, functioning like an unguarded door that allowed unrestricted access. These vulnerabilities enabled malicious users to spread malware and execute attacks with ease, perpetuating cyber threats like a persistent epidemic. To address these challenges, future network architectures must prioritize strong user authentication. Developing an advanced security framework that accurately identifies and validates all users is essential. Such a framework would ensure seamless identity integration across all network and application layers, significantly improving cybersecurity.

\subsection{The IP system is difficult to evolve}

Since its inception, the IP protocol has established itself as both the cornerstone and the standard of the Internet. However, the rapid evolution of the internet and the increasing diversity of its applications have highlighted the rigidities inherent in the IP protocol framework, which struggles to keep pace with emerging networking trends. This entrenchment poses a significant threat to the ongoing, healthy development of the Internet. Robert Kahn, one of the pioneers of the Internet, analogizes that modifying the Internet is fraught with challenges, akin to replacing the wings and engines of an airplane in flight without a place to land, which underscores the severe entrenchment of the IP system\cite{kahn}.

From a hardware perspective, the IP/TCP protocol architecture is deeply embedded within various hardware devices, thereby constraining its flexibility. Network devices, including routers and switches, typically employ firmware to implement the IP/TCP protocol stacks. This implies that updates and upgrades to the protocols often require extensive replacements and adjustments to the hardware. Such practices greatly diminish the adaptability of the network infrastructure, limiting its ability to respond quickly to emerging network needs and security threats.

From a software perspective, the vast majority of network applications and operating system kernels are predicated on the IP/TCP protocol stack, which considerably compounds the challenges of network protocol innovation. Although some emerging protocols, such as QUIC, attempt to sidestep these constraints by leveraging UDP, they still encounter substantial hurdles related to compatibility and performance optimization. This entrenchment at the software level not only restricts the scope for protocol innovation but also amplifies the complexity involved in development and maintenance, thus hindering the ecosystem's ability to respond quickly to technological advances and evolving user demands.

The challenges of transitioning from IPv4 to IPv6 are evident. As the number of network devices surges, the IPv4 address space becomes increasingly depleted. In contrast, IPv6 offers a vastly expanded address space\cite{ashraf2020analyzing}. Despite more than two decades of promotion by governments and international organizations, the adoption and implementation of IPv6 remain hampered by hardware and software compatibility issues, resulting in a coverage that does not yet exceed 50\%.

The explosive growth in video content has fundamentally altered the demands placed on network architectures, with vast amounts of network bandwidth devoted to transmitting redundant video data, thereby causing congestion across global networks. This surge in demand for efficient content distribution has led both industry and academia to re-assess IP network architectures. A content-centric network architecture offers a superior solution for the distribution of substantial video data volumes. It can significantly decrease the volume of redundant data transmitted over the network, accelerate user access to resources, alleviate the burden on the network infrastructure, and promote the sustainable development of the network\cite{de2013information}. Researchers from various countries have introduced multiple prospective network architectures; however, these future network architectures are not compatible with existing IP networks. The exclusivity inherent in IP networks impedes the deployment of these novel network architectures, thus severely limiting the diversity and evolutionary potential of global network architectures.

A critical component of advancing future network development lies in the creation of a unified architecture that ensures compatibility across diverse network frameworks while fostering innovation and diversity in network architectures. A gradual transition from the existing IP network infrastructure is essential to minimize the costs associated with system evolution. This approach not only mitigates the redundancy of infrastructure but also reduces energy consumption, thereby supporting the sustainable and inclusive growth of the internet.

\section{One solution: Co-governed Multi-Identifier Network}\label{sec3}

\begin{figure}[htbp]
    \centering
    \includegraphics[width=0.8\columnwidth]{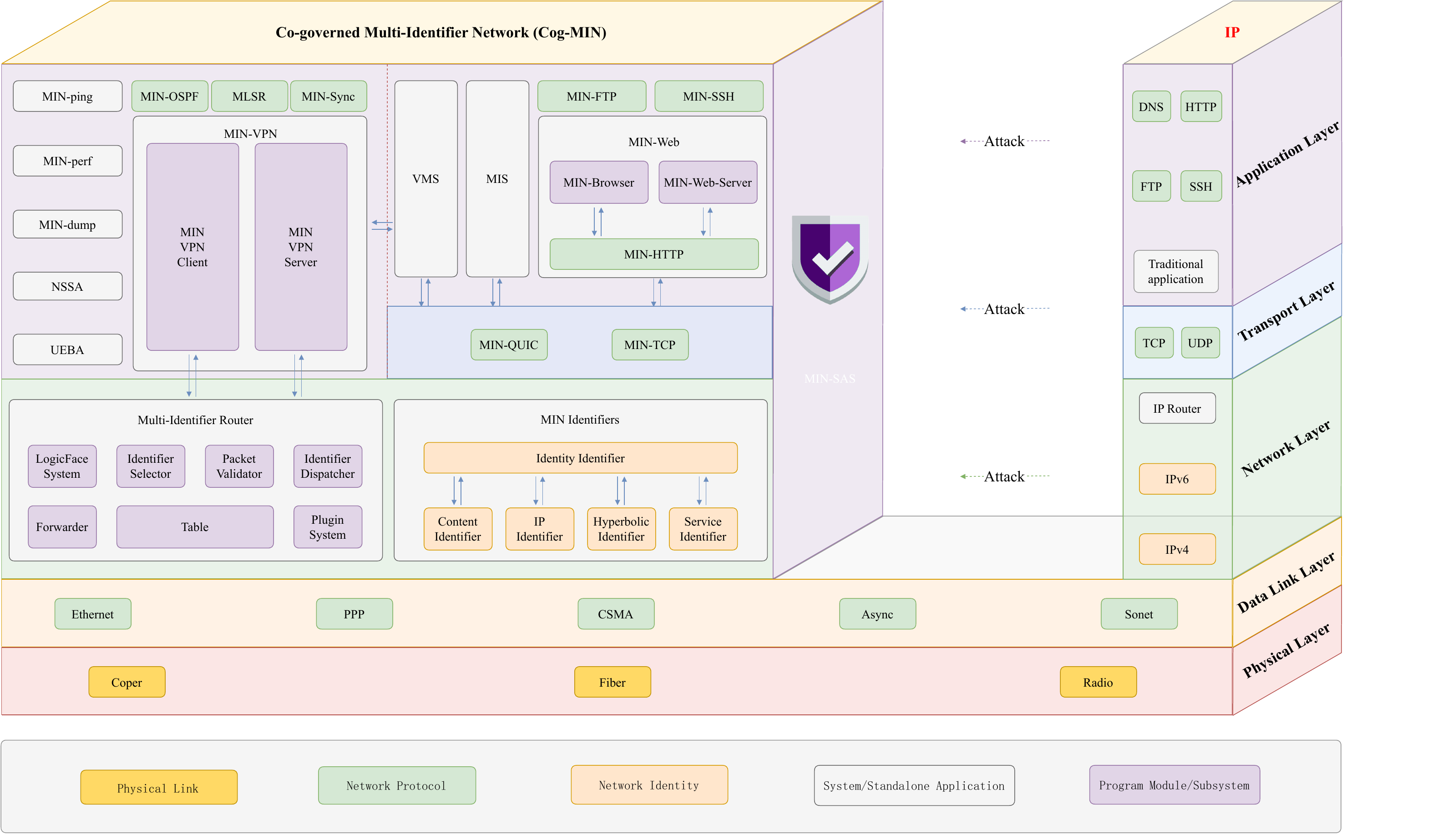}
    \caption{Overall architecture of MIN }
    \raggedright{\small  At the physical layer, MIN employs established protocols and hardware. Similarly, the data link layer leverages proven technologies reinforcing reliability and performance. Significant deviations from traditional IP systems become apparent at the network layer and above.}
    \label{fig:MINall}
\end{figure}

MIN integrates consortium blockchain technology with future network architectures while maintaining compatibility with IP, allowing for a seamless and gradual transition away from traditional IP-based systems. Its hierarchical structure, depicted in Figure~\ref{fig:MINall}, reflects a thoughtfully layered design. At the physical layer, MIN employs established protocols and hardware, such as Copper, Fiber, and Radio, to ensure efficient signal transmission. Similarly, the data link layer leverages proven technologies, including Ethernet, PPP, CSMA, Async, and Sonet, reinforcing reliability and performance.

Significant deviations from traditional IP systems become apparent at the network layer and above. Unlike conventional IP systems, MIN supports push semantics within IP networks while simultaneously accommodating pull semantics characteristic of Information-Centric Networking (ICN) \cite{que2021network}. This dual compatibility enables MIN to process network packets flexibly across diverse architectural paradigms.

By integrating these features, MIN addresses the evolving demands of modern networks, demonstrating a deep understanding of their requirements. This innovative architectural design not only enhances flexibility but also establishes a robust foundation for future network development, positioning MIN as a pivotal system in the evolution of network infrastructure.

\subsection{Basic architecture of MIN}

The network layer serves as the cornerstone of the MIN architecture, incorporating a mechanism for the expansion and evolution of multi-identifiers, which is fundamentally based on Identity Identifiers. The currently supported types of identifiers include:

\begin{enumerate}[\indent(a)]
    \item Cryptographic-based Identity Identifier
    \item Content Identifier
    \item Service Identifier
    \item IP Identifier
    \item Hyperbolic Identifier
\end{enumerate}

In addition, within the MIN system, identifiers can also include Internet of Things  devices, smart connected vehicles, mobile phone numbers, and MAC addresses.

The MIN network packet utilizes Tag-Length-Value (TLV) encoding, which enhances its flexibility and adaptability to diverse network requirements. A standard MIN network packet consists of four primary sections: the identifier area, signature area, read-only area, and variable area. These sections collectively support the efficient transmission and processing of data across the network.

\begin{figure}[htbp]
    \centering
    \includegraphics[width=0.8\columnwidth]{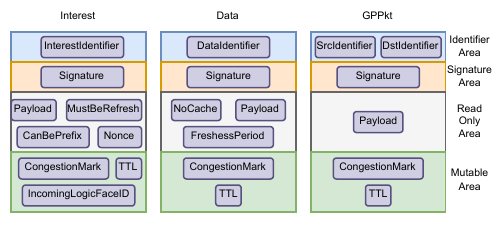}
    \caption{Different MIN network packet structures\cite{que2021network}}
    \raggedright{\small MIN network packets are classified into three distinct types: Interest Packets, Data Packets, and General Push Packets (GPPkt). Under pull semantics, users initiate requests for specific content by transmitting Interest Packets. In response, Data Packets are delivered by the content provider or by storage nodes within the network. By contrast, push semantics are implemented through GPPkt, which function similarly to the push mechanisms of traditional IP networks.  }
    \label{fig:MINpackets}
\end{figure}

MIN network packets are classified into three distinct types: Interest Packets, Data Packets, and General Push Packets (GPPkt). Under pull semantics, users initiate requests for specific content by transmitting Interest Packets. In response, Data Packets are delivered by the content provider or by storage nodes within the network, ensuring the requested data reaches the user. By contrast, push semantics are implemented through GPPkt, which function similarly to the push mechanisms of traditional IP networks. This design enables proactive data dissemination, extending the versatility of the MIN architecture. Figure~\ref{fig:MINpackets} provides a detailed visualization of the structural composition of these MIN network packets, illustrating their functionality and adaptability within diverse networking scenarios.

At the transport layer, MIN incorporates advanced designs inspired by IP networks, particularly the TCP and QUIC protocols. Building on these frameworks, the MIN-TCP and MIN-QUIC protocols \cite{min-quic} have been developed to support push-based reliable stream transmission over the MIN network layer. These protocols ensure robust end-to-end reliability within the multi-identifier networks that MIN facilitates. To address the specific requirements of pull-based content delivery, the MIN-PTCP protocol \cite{yin2022effective} has been introduced, providing comprehensive support for efficient and reliable transmissions. Notably, MIN networks redefine the traditional role of the transport layer, partially reducing its prominence. This reconfiguration allows application layer protocols and applications to operate either on top of the transport layer or directly over the network layer. This innovative architectural design enhances the flexibility and adaptability of applications within the MIN framework, enabling more efficient and tailored interactions between layers while meeting diverse network demands.

At the application layer, MIN has been adapted to support various protocols and applications:

\begin{enumerate}[\indent(a)]
    \item MIN-HTTP: HTTP transmission in MIN environment.
    \item MIN-WEB: Development tools of MIN website.
    \item MIN-VPN: It allows MIN to cross the IP network.
    \item VPN-Management-System: Implement user management and rule configuration of MIN-VPN.
    \item MIN-PING: A tool for detecting the access and communication latency of the destination.
    \item MIN-PERF: A tool for measuring throughput between MIN nodes.
    \item MIN-DUMP: A tool used for capturing and analyzing MIN network traffic.
    \item Network Security Situation Awareness: Full traffic detection is performed on incoming IP traffic to the MIN network.
    \item MIN-FT: Efficient file transfer in MIN.
    \item  MIN-SSH: Implement secure remote login sessions in MIN.
\end{enumerate}

The implementation of MIN hinges on two foundational components: the Multi-Identifier System (MIS) \cite{MIS,HMIS} and the Multi-Identifier Router (MIR) \cite{zhang2020mir} which will be introduced in Section~\ref{sec:mis} and Section~\ref{sec:mir}

\subsection{Multi-Identifier System}
\label{sec:mis}
Since its inception, the internet was envisioned as a distributed, egalitarian, and open global system. However, the centralized management of IP address allocation, top-level domain assignment, and DNS resolution has impeded the ability of nations to collaborate effectively in cyberspace governance. This centralization has given rise to unilateral monopolies, a phenomenon that starkly contrasts with the internet's original spirit of openness. Sir Tim Berners-Lee, the creator of the World Wide Web, has expressed profound dissatisfaction with this centralization, emphasizing that reclaiming the principle of decentralization is essential to addressing the internet's current challenges\cite{mansour2016demonstration}.

In recent years, blockchain technology, recognized for its decentralized and secure nature, has emerged as a transformative force, garnering widespread attention. Our proposed GBT-CHAIN and its consortium blockchain consensus algorithm, PPoV \cite{li2017proof,wang2021data}, offer an effective solution to the CAP trilemma \cite{wang2024gbt} in distributed systems. This system achieves complexity at the theoretical lower bound while attaining optimal performance in practical engineering applications. Additionally, we have pioneered the application of blockchain technology in internet management, presenting an innovative framework aimed at fostering the internet's decentralized development. This research not only expands the potential of blockchain applications but also establishes a robust foundation for the future transformation of internet architecture.

\begin{figure}[htbp]
    \centering
    \includegraphics[width=1\linewidth]{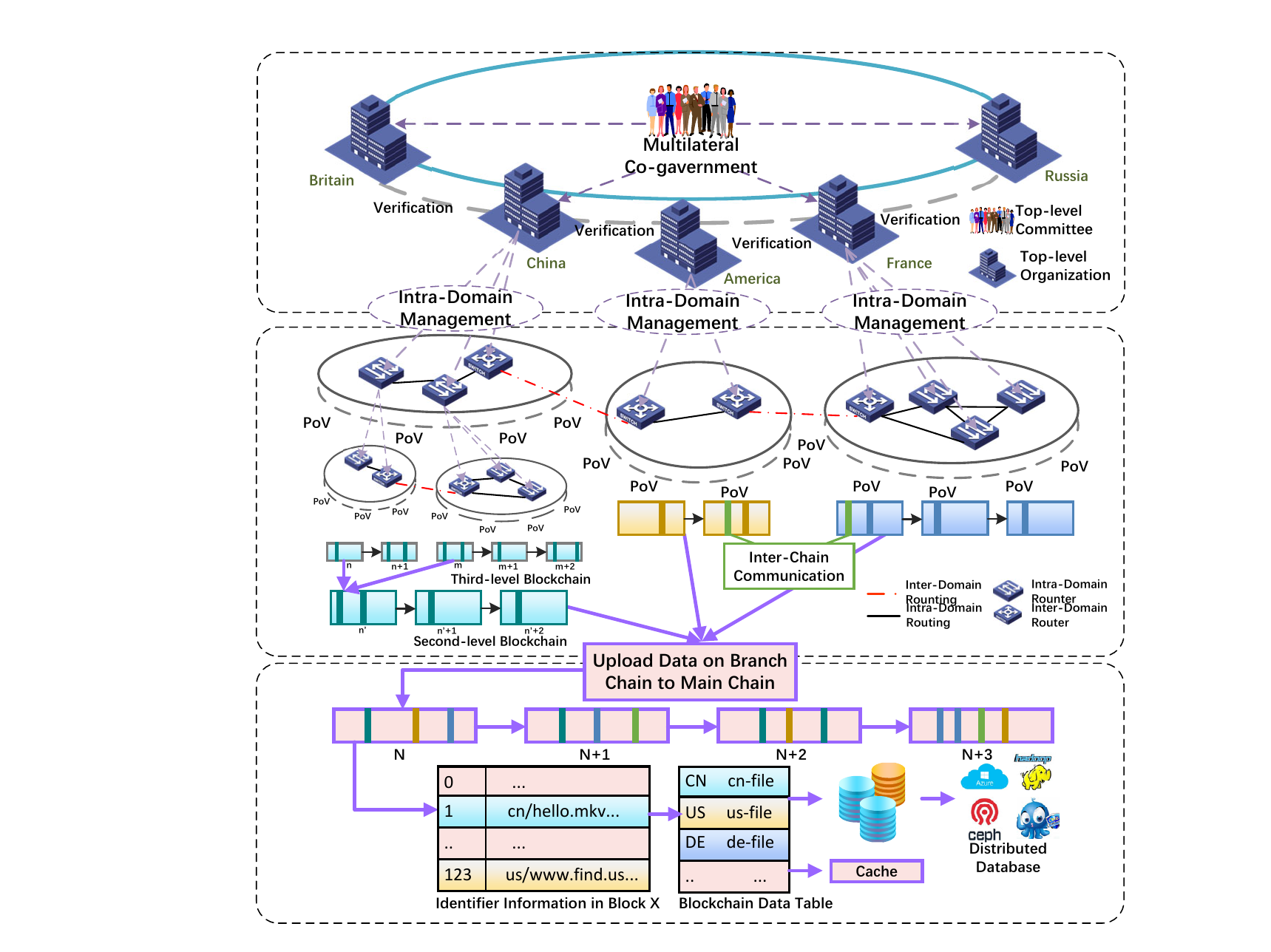}
    \caption{Multi-Identifier System achieves co-governed network\cite{li2021co}}
    \raggedright{\small The MIN cyberspace is multilaterally co-governed by a tree of blockchain. MIN divides the whole network into hierarchical domains from top to bottom. The top-level identifiers are co-governed by all nations to ensure interconnection, while the lower-level identifiers are independently managed by each nation, safeguarding sovereign independence. This makes it possible to build a secure, trustworthy and rule-based cyberspace.}
    \label{fig:mis}
\end{figure}

As depicted in Figure~\ref{fig:mis}, we have systematically designed a blockchain-based multi-identifier management system that supports functions such as the registration, resolution, and mutual translation of various identifiers \cite{MIS,HMIS}. This system is structured into two hierarchical levels: the consortium blockchain management level and the consortium blockchain control level. At the management level, a top-level identifier management committee, composed of governments worldwide, governs identifier permissions and functionalities through a transparent voting mechanism. This approach facilitates collaborative governance of the global network, enhancing both transparency and democracy in the management of cyberspace \cite{wei2021co}. By addressing the structural limitations of centralized systems, this innovative model represents a significant step toward a decentralized and inclusive internet.

In the process of routing and forwarding, the MIN architecture incorporates a consortium blockchain-based Multi-Identifier management system, seamlessly integrated at the consortium blockchain control layer. This system employs an efficient consensus algorithm to achieve agreement among nodes while recording routing states and request authentication information within the domain. At the routing layer, the system resolves multiple types of identifiers, including address identifiers, content identifiers, and identity identifiers. To ensure data integrity and immutability, up-chain and down-chain data undergo hashing for verification. This blockchain-based verification strategy effectively addresses the bottleneck associated with large-scale data queries on the blockchain, significantly enhancing the service efficiency of the Multi-Identifier System.

In an experiment involving 200 server nodes distributed across 26 countries on five continents, the Multi-Identifier System demonstrated its capability to dynamically and efficiently manage and validate identifiers at multiple levels. This functionality ensures the security and traceability of top-level identifiers, reinforcing the system's robustness. Such advancements lay a critical foundation for the decentralization of global cyberspace, fostering a more secure, transparent, and collaborative network environment \cite{MIS}.

\subsection{Multi-Identifier Router}
\label{sec:mir}
The MIR serves as a pivotal element, responsible for parsing the identifiers within MIN network packets, validating their authenticity, and determining the optimal forwarding path.

The internal structure of the MIR comprises three essential tables: the Forwarding Information Base (FIB), the Pending Interest Table (PIT), and the Content Store (CS). The FIB maintains forwarding port information for Interest Packets and GPPkts, ensuring efficient and expedient delivery of packets to their destinations. The PIT tracks all forwarded Interest Packets, providing precise routing guidance for their corresponding Data Packets upon return. Meanwhile, the CS stores packets containing frequently accessed content, significantly improving access efficiency by reducing redundant requests.

\begin{figure}[htbp]
    \centering
    \includegraphics[width=0.8\columnwidth]{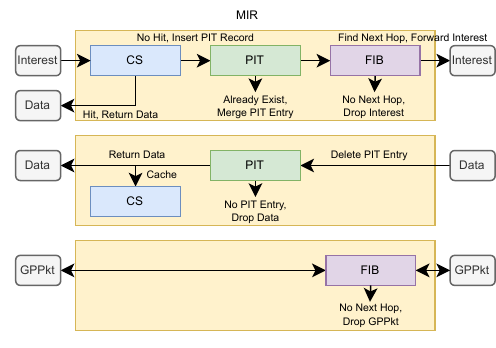}
    \caption{Processing flow of MIR for different MIN network packets\cite{que2021network}}
    \raggedright{\small The MIR mainly contains three tables, Forwarding Information Base(FIB), Pending Interest Table(PIT), and Content Store(CS). The FIB stores the port that Interest Packet and GPPkt should forward to. PIT records all Interest packets that have been forwarded, which are used to guide returned Data Packets. CS caches the Data Packets of some popular content. }
    \label{fig:MIR}
\end{figure}

As illustrated in Figure~\ref{fig:MIR}, the processing flow of MIR for various MIN network packets showcases its critical role and high efficiency. By integrating these components, the MIR ensures seamless operation within the MIN framework, optimizing packet routing and enhancing overall network performance.

\subsection{Cyberspace Customs protects the sovereignty of countries' cyberspace borders}

In today's IP-based network environment, data packets traverse the global network freely, creating significant challenges for countries in tracing and holding foreign attackers accountable for malicious activities originating beyond their borders. While most nations focus on regulating their domestic networks, the lack of international standards and enforcement mechanisms, coupled with disparities in legal frameworks, creates opportunities for transnational cyberattacks and cybercrime. Manuel Castells, a former Spanish Minister of Universities, has observed that the transnational nature of the web undermines traditional nation-state sovereignty. Consequently, managing and asserting national sovereignty in cyberspace has emerged as a pressing and widely debated issue.

In the physical world, passports and visas play a vital role in regulating cross-border movements. A passport is an identification document issued by a country to its citizens for international travel, while a visa is a permit granted by a destination country to foreign travelers based on its laws and regulations. Governments utilize these tools to identify and monitor high-risk individuals, combat transnational crime and terrorism, and protect national borders. The application process for visas, which involves the use of passports, facilitates effective border management.

\begin{figure}[htbp]
    \centering
    \includegraphics[width=1\linewidth]{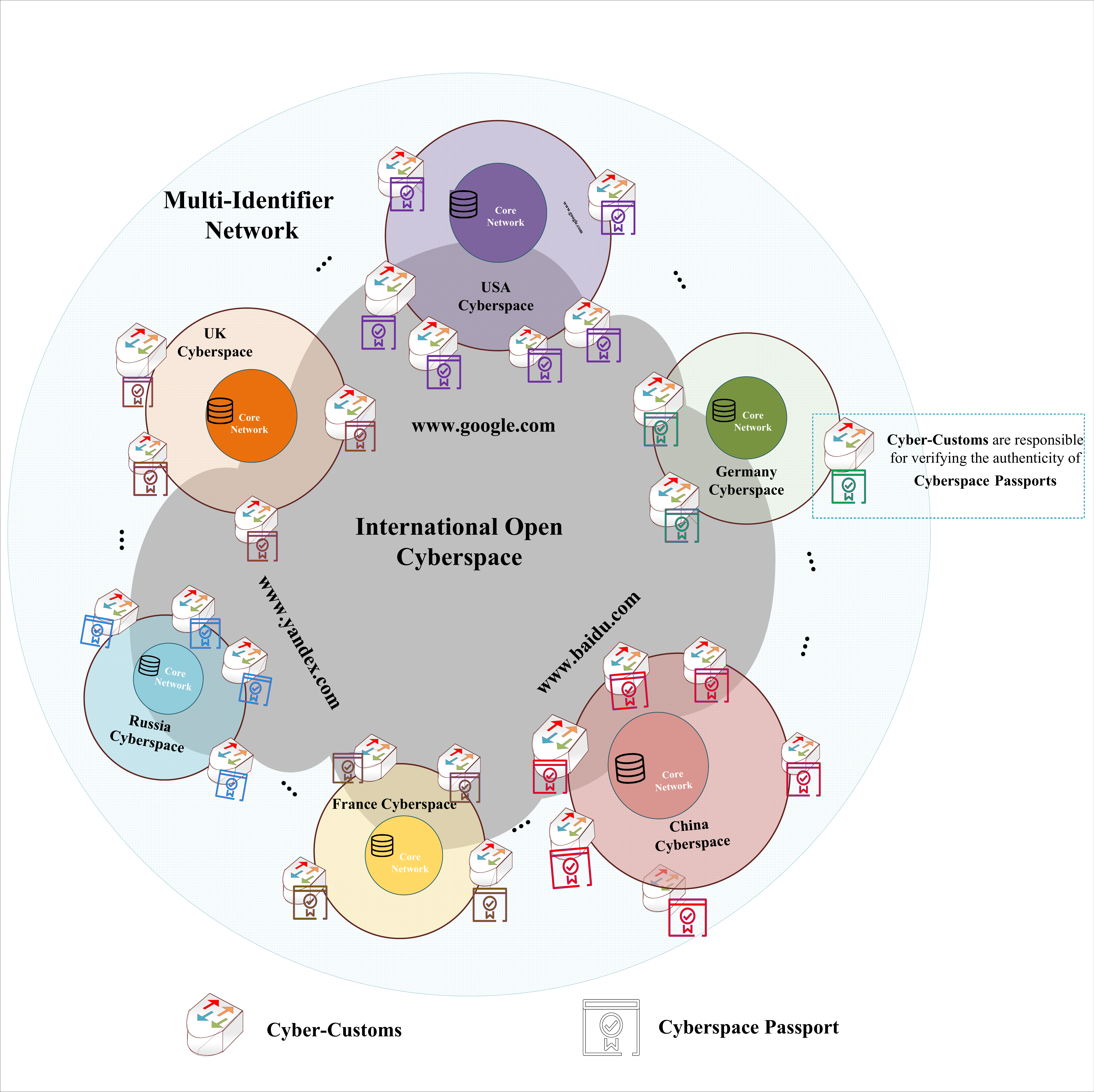}
    \caption{Cyberspace Customs protects the sovereignty of each country's cyberspace border}
    \raggedright{\small Cyberspace Passports and visas 
 shape cyberspace sovereignty. Cyberspace customs sign cyberspace passports for all MIN packets bound for international open networks. Only MIN packets with legal cyberspace visas can enter the network border from the international open cyberspace.} 
    \label{fig:custom}
\end{figure}

Drawing inspiration from these systems, we have designed an innovative border management protocol at the network layer of the Multi-Identifier Network (MIN), aimed at enhancing the governance of cyberspace boundaries. This protocol enables independent governance of network domains and facilitates the regulation and monitoring of cross-domain data flows. As illustrated in Figure ~\ref{fig:custom}, the protocol requires outgoing packets to be signed by both the user and the domain border router using a dual hash value mechanism, referred to as a cyberspace passport and a cyberspace visa. Upon arrival, the receiving domain's boundary router verifies these hash values to determine whether the packet should be accepted.

To implement this system, a user from Country A must first apply for a Cyberspace Visa Key (CVK) from the customs authority of Country B. This application uses the user's real identity and biometric information to generate the cyberspace visa hash value. Simultaneously, the customs authorities of both countries agree upon a Cyberspace Passport Key (CPK) to mutually verify each other's cyberspace passports. As shown in Eq~\ref{eq:visa}, the user calculates a cyberspace visa using Time256 and the CVK, embedding the resulting CyberVisa field into their network packet.

This protocol not only strengthens the ability of nations to manage their cyberspace boundaries but also ensures that cross-domain data flows can be traced and regulated effectively. By adopting mechanisms inspired by physical-world border management, the protocol addresses the complexities of transnational cyber governance and paves the way for a more secure and orderly cyberspace.

\begin{equation}
\label{eq:time}
\mathrm{Time} = \mathrm{UNIX~Time}~ and~\mathrm{0xFFFFFFFFFFFFFFF0}
\end{equation}

\begin{equation}
\mathrm{Time256} = \mathrm{Time} \times (1 + 2^{64} + 2^{128} + 2^{192})
\end{equation}

\begin{equation}
\label{eq:visa}
\mathrm{Visa} = \mathrm{SHA256}(\mathrm{Time256}~xor~\mathrm{CVK})
\end{equation}

As described in Eq.~\ref{eq:pass}, the Cyberspace Customs calculates the Cyberspace Passport by combining the user's Cyberspace Visa with the Cyberspace Passport Key (CPK). This calculated value is then inserted into the CyberPass field of the network packet, effectively endorsing the outbound behavior of the national user's packet. When the receiving domain verifies both the Cyberspace Visa and Cyberspace Passport, the packet is granted access to the foreign network.

\begin{equation}
\label{eq:pass}
\mathrm{Pass} = \mathrm{SHA256}(\mathrm{CPK}~xor~\mathrm{Visa})
\end{equation}

Experimental results \cite{cyberpass} demonstrate that this protocol effectively manages network boundaries, efficiently handling tens of millions of user traffic without introducing significant latency. Additionally, the protocol is resistant to attacks targeting layers above the network layer, including replay attacks and other similar threats, thereby significantly enhancing security at the network layer.

An added advantage of this system is its ability to revoke the Cyberspace Visa of malicious foreign users in real-time, providing an effective mechanism to counter transnational cyberattacks. This capability ensures robust protection of national network boundaries, offering a practical and scalable solution for enhancing the security and governance of cyberspace.

\subsection{Traceability makes hackers nowhere to hide}

The IP network protocol prioritizes the efficiency and reliability of data transmission over strict user authentication, making it inherently difficult to trace the origins of network packets. As a result, network security strategies typically focus on mitigating malicious behaviors, such as traffic monitoring, anomaly detection, and intrusion prevention, rather than identifying and addressing malicious users directly. The root cause of this challenge lies in the ease with which attackers can forge IP addresses. By employing techniques like proxy servers, attackers can easily mask their real identities, complicating traceability. Although existing security measures are essential for defending against attacks, they fall short when it comes to pinpointing the origins of such malicious activities.

Bruce Schneier, in his book Secrets and Lies, emphasizes that cyberattacks often stem from the motives and intentions of malicious users rather than solely from technical vulnerabilities \cite{schneier2015secrets}. He argues that without real-name authentication for network users, the IP system remains ill-equipped to address the fundamental issues of network security. He likens this to a fragile patient who cannot escape illness despite protective measures. To address these challenges, future network architectures must incorporate real-name authentication at a technical level, significantly reducing the risks of cyberattacks.

To overcome these limitations, the MIN architecture registers real personal information for each user within the MIS consortium blockchain, assigning them a unique cryptographic identity. This system is supported by a robust user identity management mechanism \cite{lin2022identity}. Under this framework, every packet transmitted by a user is signed with their private key, ensuring authentication during both transmission and reception. This mechanism strengthens the association between the user and their data while facilitating the accurate tracking of malicious behavior. When malicious activity is detected, offending users can be quickly and accurately identified, allowing for effective countermeasures at minimal cost and ensuring the stability and security of the network.

The widespread prevalence of cybercrime today is further exacerbated by the lack of effective legal and technical deterrents. This issue stems from two primary factors: the limitations of current legal frameworks and the deficiencies in IP-based network technology. The IP architecture lacks a robust authentication mechanism, enabling criminals to obscure their tracks with ease through technologies like VPNs, proxy servers, and the Tor network. Such methods make it challenging for law enforcement agencies to trace perpetrators, while blocking IP addresses proves to be a temporary and ineffective solution, as attackers can swiftly switch devices or accounts to continue their activities.

The MIN network addresses these challenges by integrating digital identities with biometric technology. This innovation allows for the enforcement of permanent access bans on individuals involved in serious cybercrimes. If a user's biometric data is flagged for engaging in criminal activities, they can be blocked completely from accessing the network. This capability creates a strong deterrent, as the consequences of cybercrime become far more severe and harder to evade. By combining biometric identification with cryptographic user identities, MIN offers a level of security that surpasses IP, laying the groundwork for a safer and more accountable cyberspace.

\subsection{Identifier evolution mechanism}

From the 1970s to the 1990s, network technology was in its early developmental stages. During this period, research institutions and enterprises worldwide actively explored ways to build efficient and reliable network communication systems. For instance, the Asynchronous Transfer Mode (ATM) network architecture was proposed in Europe during this time \cite{kim1995atm}. ATM is a connection-oriented high-speed data transmission technology capable of supporting various types of data transmission, including voice, video, and data. By the early 1990s, ATM networks were widely adopted in the telecommunications and financial services industries.

In addition to ATM, there are other network architectures such as X.25\cite{barrett1991lan} and Frame Relay\cite{grossman1991overview}, which are used in specific areas and regions. Each of these architectures has its advantages: X.25 is known for its reliable data transfer service, while frame relay is favored for its lower cost and higher flexibility.

However, with the widespread adoption of the Internet Protocol (IP), particularly the Transmission Control Protocol/Internet Protocol (TCP/IP) model, IP networks gradually gained dominance. This dominance posed significant challenges to other network architectures, many of which were gradually phased out due to their inability to compete with IP networks. The trend toward global network homogenization has become increasingly evident, limiting technological diversity and potentially threatening the security and stability of networks. In his book \cite{carr2020shallows}, renowned American author Nicholas G. Carr expressed concerns about the vulnerability of the internet's infrastructure. When all networks rely on a single architecture, the consequences of a security breach or technical defect can be profound and catastrophic.

In addition, IP network packets are designed with fixed fields and formats, preventing the packet structure from being updated to meet new requirements arising from network evolution. In the IP system, routers are limited to supporting IP addresses as the basis for forwarding. To adapt to changing network scenarios, we have witnessed significant changes in the packet formats and addressing schemes of IPv4 and IPv6. However, a large number of network devices have been forced into obsolescence due to their inability to support IPv6 packet formats and addressing methods, resulting in substantial economic burdens, a significant accumulation of electronic waste, and a considerable slowdown in the network upgrade process.

\begin{figure}[htbp]
    \centering
    \includegraphics[width=0.8\linewidth]{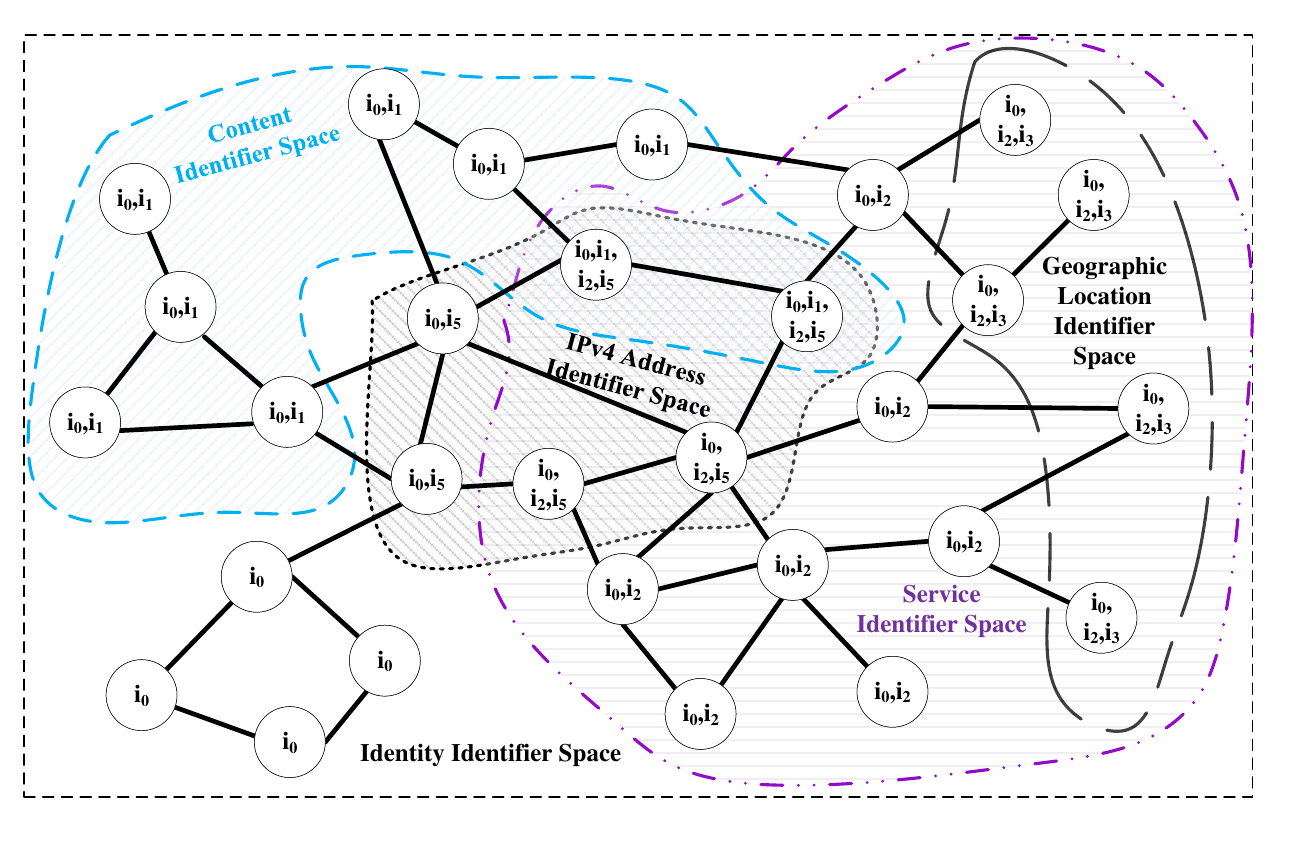}
    \caption{Different network architectures can coexist in MIN\cite{MIS}}
    \raggedright{\small The figure illustrates graphically how the identifier space is divided in the network. Each circle indicates the node and its identifier types. Identity is the basic identifier that all nodes own, so the entire network is the identity identifier space.}

    \label{fig:multi}
\end{figure}

The network packet of MIN is encoded in TLV format, enabling its structure to be continuously updated and evolved. Furthermore, MIN supports both Push and Pull semantics, ensuring compatibility with various ICN architectures. A scalable communication scheme at the network layer has also been proposed in MIN, incorporating a scalable multi-identifier management scheme and a universal addressing mechanism based on a scalable identity architecture \cite{wang2021scalable}. MIN introduces an extended identifier management solution that supports the coexistence of multiple identifiers. By using encrypted identifiers, MIN simultaneously maintains compatibility with various identifier types, such as IP addresses. The MIR provides a general routing scheme that includes an identifier detection mechanism and a candidate identifier sorting algorithm. By analyzing relationships between multiple identifiers as a graph, MIR selects valid identifiers for routing. As shown in Figure~\ref{fig:multi}, MIN facilitates the continuous expansion and evolution of identifier types and communication modes at the network layer, ensuring compatibility with diverse network identifiers \cite{enl}.

As a result, MIN supports the coexistence of different network architectures. When deployed as a network architecture, MIN allows existing IP network devices to remain operational, enabling users to continue using the IP protocol. IP network packets can be transmitted as usual within an IP network. When these packets reach an MIR running the MIN protocol, the MIR routes the IP packets using a valid identifier based on the general identifier routing scheme. This framework provides a low-cost and sustainable solution for network upgrades. Upgrading a network architecture in the IP system can be likened to scaling a steep cliff, whereas upgrading within the MIN system feels akin to walking effortlessly along mountain trails. MIN enables the IP system to transition seamlessly and gradually to future network architectures, eliminating many of the challenges traditionally associated with such upgrades.

In addition, MIN supports the use of distinct network architectures by different countries or organizations. As long as these architectures employ TLV-encoded network packets and register their identifiers in the MIS, the MIR can select a suitable identifier for routing. In this scenario, diverse network architectures from various countries or organizations can coexist within the MIN ecosystem. This capability enhances the robustness and diversity of global network architectures, restoring the rich variety that characterized the early internet. It also promotes the sustainable development of network systems, fostering a resilient and inclusive future for global communications.

\subsection{Multi-Identifier Network with quantum information technology}
\begin{figure}
    \centering
    \includegraphics[width=0.5\linewidth]{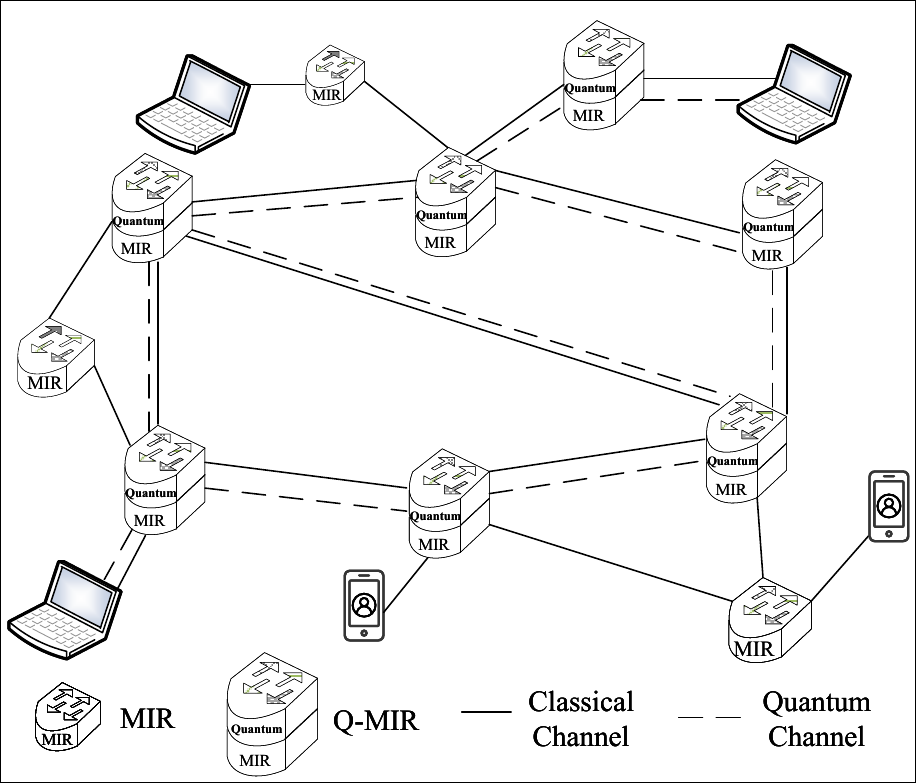}
    \caption{Deployment of Q-MIN with classical and quantum dual-channel\cite{li_CAP_2024}}
    \raggedright{\small In the gradual realization of the quantum internet, one exploratory approach is to integrate quantum technologies with existing network frameworks. The Multi-Identifier Network (MIN), benefiting from its diverse types of identifiers and excellent scalability, offers the potential to incorporate quantum identifiers into the system, forming a Multi-Identifier Network that supports quantum identifiers (Q-MIN). This integration further enhances the security of the MIN and improves its competitiveness in the quantum era, serving as both an exploration and an attempt at future quantum network communications. Since the MIN framework is derived from the evolution of classical network systems, and there are fundamental differences in the physical implementation of classical and quantum communications, the two cannot be fully compatible. Therefore, dual-channel support is required. The figure illustrates a deployment strategy for Q-MIN, where quantum hardware and software are deployed alongside classical hardware and software, working in tandem without interfering with each other.} 
    \label{fig:qit}
\end{figure}

Quantum information technology can revolutionize the security of the MIN system. Quantum Key Distribution (QKD) leverages the principles of quantum mechanics to achieve unconditional security in information transmission. As Figure~\ref{fig:qit} shows, the MIN system can adopt QKD to establish encrypted communication channels, ensuring data transmission is immune to eavesdropping and tampering, thereby fundamentally enhancing network security\cite{patent}. By integrating blockchain technology, the Multi-Identifier System (MIS) in MIN can simultaneously ensure communication privacy and enhance the credibility of user authentication.

As quantum computing advances, traditional encryption algorithms may face vulnerabilities. The MIN system adopts quantum-resistant cryptographic methods, such as lattice-based or hash-based encryption techniques, to address future quantum computing challenges and ensure the security and sustainability of the network architecture in the quantum era.

MIN can utilize quantum random number generators to produce truly unpredictable random numbers for key generation and other high-security operations\cite{hanfeng}. This approach offers greater security than pseudo-random numbers generated by traditional algorithms, further strengthening the system's defenses.

Blockchain technology forms the core of the MIS. By employing quantum computing-optimized on-chain verification mechanisms, data processing security and transaction verification fairness can be significantly improved. Additionally, quantum computing can rapidly solve complex graph-matching problems, further optimizing the management and coordination of multiple identifiers within MIS\cite{li_CAP_2024,lin2024Q-PnV}.

The flexibility and adaptability of the MIN architecture enable seamless integration with future quantum internet technologies. By incorporating quantum network nodes, MIN can facilitate efficient collaboration between quantum computing nodes, laying a solid foundation for the diversification and sustainable development of the global network ecosystem.

\section{A wide range of prototype testing and anti-attack security testing of MIN}

This section begins by addressing the severe security challenges currently faced by the Internet of Things (IoT), emphasizing the significant threats these challenges pose to human safety. It then details the foundational deployment and security testing of MIN, followed by a demonstration of the practical reliability of MIN-VPN and MIN-V2X in top-tier international security competitions, underscoring the high level of security that MIN provides.

\subsection{IoT security is alarming}

With the rapid development of the Internet of Things, cyberattacks have evolved beyond software exploitation, directly endangering human safety. A particularly alarming incident occurred in 2015 when hackers exploited vulnerabilities in a car's IP network, gaining access to and rewriting the firmware of specific vehicle chips. Through the In-Vehicle Network Architecture, they issued commands to disable the brakes and take control of the steering wheel, ultimately forcing the car into a ditch. This security defect affected 1.4 million vehicles produced by Fiat Chrysler Automobiles, prompting a massive recall of all affected vehicles \cite{eiza2017driving}.

Even advanced vehicles like Tesla are not immune to cyber threats. In 2022, a 19-year-old hacker remotely compromised over 25 Tesla vehicles across 13 countries. Exploiting the vehicles' network systems, the hacker was able to open doors and windows, control radios and headlights, and even enable “keyless driving” by starting the engines remotely \cite{tesla}. These incidents highlight the grave security vulnerabilities in intelligent connected vehicles, raising significant concerns across the industry.

The seriousness of these threats was further underscored during the 2021 International Elite Challenge on Cyber Mimic Defense, which introduced a connected vehicle track for the first time \cite{wu2020application}. Leading cybersecurity teams from countries including Russia, China, Ukraine, India, Japan, Singapore, Malaysia, and Israel participated in the event. Over the course of three hours, hackers successfully manipulated 15 vehicles from prominent international brands based on IP networks. These vehicles, from manufacturers in the United States, Germany, and China, were compromised through their Controller Area Network (CAN) systems. The results were profoundly alarming: Successful exploitation during vehicle operation could lead to catastrophic consequences, including severe casualties and significant property damage.

This growing landscape of IoT security threats demonstrates the critical need for robust, innovative solutions. Addressing these vulnerabilities is essential not only for protecting data, but also for safeguarding human lives in increasingly connected environments.

\subsection{MIN prototype deployment and security testing}

To evaluate the feasibility of MIN for large-scale deployment, we collaborated with the institutions listed in Table~\ref{tab:test} to implement and test MIN. The experimental results demonstrated that MIN's rationality testing in large-scale scenarios successfully met its design requirements \cite{li2021co,9000531}.

\begin{table}[htbp]
  \centering
  \caption{MIN large-scale deployment test}
    \begin{tabular}{cc}
    \toprule
    Institution & Type \\
    \midrule
    China Telecom Futian & Telecommunications carrier \\
    Peking University Shenzhen Graduate School & University \\
    China Unicom Tongle & Telecommunications carrier \\
    GuangDong Communications \& Networks Institue & Corporation \\
    Kingsoft & Corporation \\
    China Unicom yuandong & Telecommunications carrier \\
    Southern University of Science and Technology & University \\
    Guangdong University of Technology & University \\
    The Hong Kong University of Science and Technology & University \\
    Macau University of Science and Technology & University \\
    The Chinese University of Hong Kong & University \\
    China Telecom Nanshan & Telecommunications carrier \\
    \bottomrule
    \end{tabular}%
  \label{tab:test}%
\end{table}%

To systematically validate the safety and security of MIN, we engaged several professional security teams to conduct comprehensive testing. Based on MIN, we developed MIN-VPN \cite{min-v2x2}, which was deployed on a carrier's IP network. Using mainstream IP-VPN as a reference system, these teams utilized classical methodologies, attack chains, and advanced security tools for comparative attack testing.

After extensive and prolonged testing by these independent professional teams, the results demonstrated that MIN-VPN effectively resisted traditional network attacks across all stages of the attack chain, regardless of whether the testing scenario involved IP-MIN or MIN-MIN networks. In contrast, IP-VPN consistently failed to provide the same level of protection. The detailed test results, as summarized in Table~\ref{tab:sectest}, clearly highlight the superior security performance of MIN-VPN compared to its traditional IP-based counterpart.

\begin{table}[htbp]
  \centering
  \caption{Professional Security Test on MIN}
  
    \resizebox{\textwidth}{!}{
    \begin{tabular}{|c|c|c|c|c|c|}
    \toprule
    \textbf{Attack phase} & \multicolumn{2}{c|}{\textbf{Attack means}} & \textbf{IP-IP test result} & \textbf{IP-MIN test result} & \textbf{MIN-MIN test result} \\
    \midrule
    \multirow{4}[8]{*}{Reconnaissance} & \multicolumn{2}{c|}{Host discovery} & Can be discovered & Discover hosts that are not target's & Nothing found \\
\cmidrule{2-6}          & \multicolumn{2}{c|}{Ping Scan} & Can be detected & Undetectable & Undetectable \\
\cmidrule{2-6}          & \multicolumn{2}{c|}{Operating system identification} & System fingerprint can be obtained & Nothing found & Nothing found \\
\cmidrule{2-6}          & \multicolumn{2}{c|}{Port scan} & Can detect all port services & Can't find any port information & Nothing found \\
    \midrule
    \multirow{4}[8]{*}{Exploitation} & \multirow{3}[6]{*}{Trojan implantation} & TCP trojan & Connected & Can't connect & Can't connect \\
\cmidrule{3-6}          &       & UDP trojan & Connected & Can't connect & Can't connect \\
\cmidrule{3-6}          &       & ICMP trojan & Connected & Can't connect & Can't connect \\
\cmidrule{2-6}          & \multicolumn{2}{c|}{Web shell} & Connected & Can't connect & Can't connect \\
    \midrule
    \multirow{2}[4]{*}{Actions} & \multirow{2}[4]{*}{ARP poisoning} & Sniff & Sniff successfully & Non-intranet & Sniff Failed \\
\cmidrule{3-6}          &       & Forced disconnection & Disconnected & Non-intranet & Failed \\
    \bottomrule
    \end{tabular}}%
  \label{tab:sectest}%
\end{table}

\subsection{MIN-VPN practical security test}

At the 2021 International Elite Challenge on Cyber Mimic Defense, we provided an attack range based on the MIN architecture for participating hacker teams. Each team was given a basic account and password to access the network, allowing them to attempt penetration attacks. Over 72 hours, the MIN architecture underwent high-intensity attack testing from 48 professional hacker teams worldwide. During this period, the system faced nearly 5,000 attacks, with over 100 malicious files uploaded to the server. Despite this intense effort, no hacker team succeeded in breaching the system, and the bounty offered remained unclaimed, showcasing the architecture's robust security.

Further validation was provided by the ISCC International Information Security Competition, organized by the former Ministry of Military Engineering of China and hosted by the Beijing Institute of Technology. Over the course of 25 days, the competition recorded a staggering 10,417,598 attacks, with attack traffic totaling 34.14 GB. Attackers uploaded 3,156 files, including 496 malicious files, and the attacks originated from 1,725 distinct IP addresses. Despite the sheer scale and intensity of these attacks, no hacker team managed to compromise the MIN-V2X system, leaving the bounty unclaimed. These extensive tests verified the robustness and reliability of MIN-based network systems, demonstrating their ability to withstand real-world cyberattacks. The detailed results of MIN's performance as a security testing platform are summarized in Table~\ref{tab:minvpn}, further solidifying its reputation as a highly secure network architecture.

\begin{table}[htbp] 
\centering
  
  \caption{MIN-VPN practical test}
    \resizebox{\textwidth}{!}{
    \begin{tabular}{ccccc}
    \toprule
    Year    & Name   & Type & Number of Source & Number of Attack \\
    \midrule
    2021  & Information Security and Countermeasures Contest & Unselected teams & 1725  & 10417598 \\
    2021  & International Elite Challenge On Cyber Mimic Defense & Selected teams &48    & 4934 \\
    2022  & Information Security and Countermeasures Contest & Unselected teams & 554   & 14713579 \\
    2022  & Gaona Aero Material Co.,Ltd & Selected teams & 10    & 387 \\
    2022  & International Elite Challenge On Cyber Mimic Defense & Selected teams & 20    & 1206 \\
    
    \bottomrule
    \end{tabular}}
    
  \label{tab:minvpn}%
  
\end{table}%

\subsection{MIN-V2X practical security test}

Based on the MIN architecture, we developed the MIN-V2X architecture \cite{min-V2X,min-v2x2} specifically for vehicle networking applications. In collaboration with leading automotive companies, we replaced traditional IP networks with MIN and implemented the MIN-V2X architecture in real vehicles. This enabled vehicle owners to remotely control critical functionalities such as steering and movement, leveraging the advanced capabilities of MIN.

To rigorously test the security of MIN-V2X, vehicles equipped with this architecture were used as “shooting ranges” in seven authoritative security competitions, as listed in Table~\ref{tab:minv2x}. Hacker teams were provided with basic account credentials to log into our servers, allowing them to attempt attacks on the system. Top-tier hacking teams from renowned organizations, including Xiaomi Tech, Yu An Security Tech, Huazhong University of Science and Technology, NSFOCUS Tech, TOPSEC Tech, and Gogobyte Tech, participated in these competitions. Despite the expertise of these teams, none succeeded in taking control of the vehicles, and no bounties were claimed.

Unlike traditional Capture The Flag (CTF) competitions, these tests were conducted in realistic, real-world scenarios, involving actual vehicles and operational networks. The results unequivocally demonstrated that MIN-V2X is significantly more secure than traditional IP-based networks, withstanding advanced and persistent attacks from some of the world's top hacking teams. This exceptional safety performance underscores MIN-V2X as the safest and most reliable choice for the future architecture of connected vehicles, setting a new benchmark for security in vehicle networking systems.

\begin{table}[htbp]
  \centering
  \caption{MIN-V2X practical test}
  \resizebox{\textwidth}{!}{
    \begin{tabular}{ccccc}
    \toprule
    Year    & Name  & Type  & Number of Source & Number of Attack \\
    \midrule
    2023  & Information Security and Countermeasures Contest & Unselected teams & 41993 & 23458616 \\
    2023  & World Intelligent Driving Challenge & Selected teams & 30    & 3998 \\
    2023  & National Intelligent Driving Test Competition of China & Selected teams & 29    & 7011 \\
    2023  & China Automobile Vulnerability Database & Selected teams & 9     & 241 \\
    2023  & International Elite Challenge On Cyber Mimic Defense & Selected teams & 202   & 6431 \\
    2024 & Information Security and Countermeasures Contest & Unselected teams & 422  & 12958432 \\
    2024 & Foshan QiYuan Industrial Internet Security Competition & Selected teams & 16  & 246 \\  
    \bottomrule
    \end{tabular}}
  \label{tab:minv2x}%
\end{table}%

\subsection{Industrial applications of MIN}

In October 2023, nearly 30 top autonomous vehicle companies set two group standards that use MIN as the communication standard for low-speed autonomous vehicles. MIN guarantees the security of autonomous vehicles and provides reliable solutions for the industry. MIN is not only an innovation in research but also an application in industry. 

At present, two of the largest telecom operators in China have adopted MIN as their business technology solutions. One uses MIN's high-security features to provide MIN private network services without IP. Another uses MIN's traceable characteristics for managing its cross-border network operations, highlighting the potential of the architecture to address complex networking needs in diverse and large-scale environments. Telecom operators participate in MIN's development, making MIN more perfect. With the promotion of telecom operators, MIN will gradually become the standard of China's communication.

\section{Cyber-security Theorems \& Conjecture Trilogy}
In this section, we propose a trilogy of theorems and conjecture addressing cyberspace security. We establish that no technical solution exists within the IP framework to ensure deterministic cyberspace security. However, using IP space as a reference, we demonstrate the potential for future architectures to provide deterministic security solutions, exponentially enhancing resilience against network attacks. Finally, we conjecture the possibility of achieving deterministic security in future cyberspace. Together, this trilogy outlines a comprehensive path for advancing cyberspace security.

\subsection{Theorem A: There is no technical solution for IP systems to guarantee deterministic cyberspace security}
This theorem is proven using proof by contradiction and enumeration:
\begin{enumerate}[\indent(a)]
    \item \textbf{Assumption}: If a deterministic security solution S1 exists for IP systems, it should already be implemented by the United States, the originator of IP technology and current manager of cyberspace.
    \item \textbf{Reality}: Despite decades of leadership in cybersecurity and the strategic advantage of managing the DNS, the United States remains a frequent victim of major cybersecurity incidents, consistently appearing among the top 10 annually.
    \item \textbf{Conclusion}: If the United States cannot develop  S1  despite its resources and expertise, the probability of any other nation achieving it within the IP framework is effectively zero.
\end{enumerate}

Therefore, there is no technical solution within IP systems to guarantee deterministic cyberspace security.

\subsection{Theorem B: Future cyberspace architectures can provide security solutions with exponential performance improvements}

\begin{figure}[htbp]
    \centering
    \includegraphics[width=0.7\linewidth]{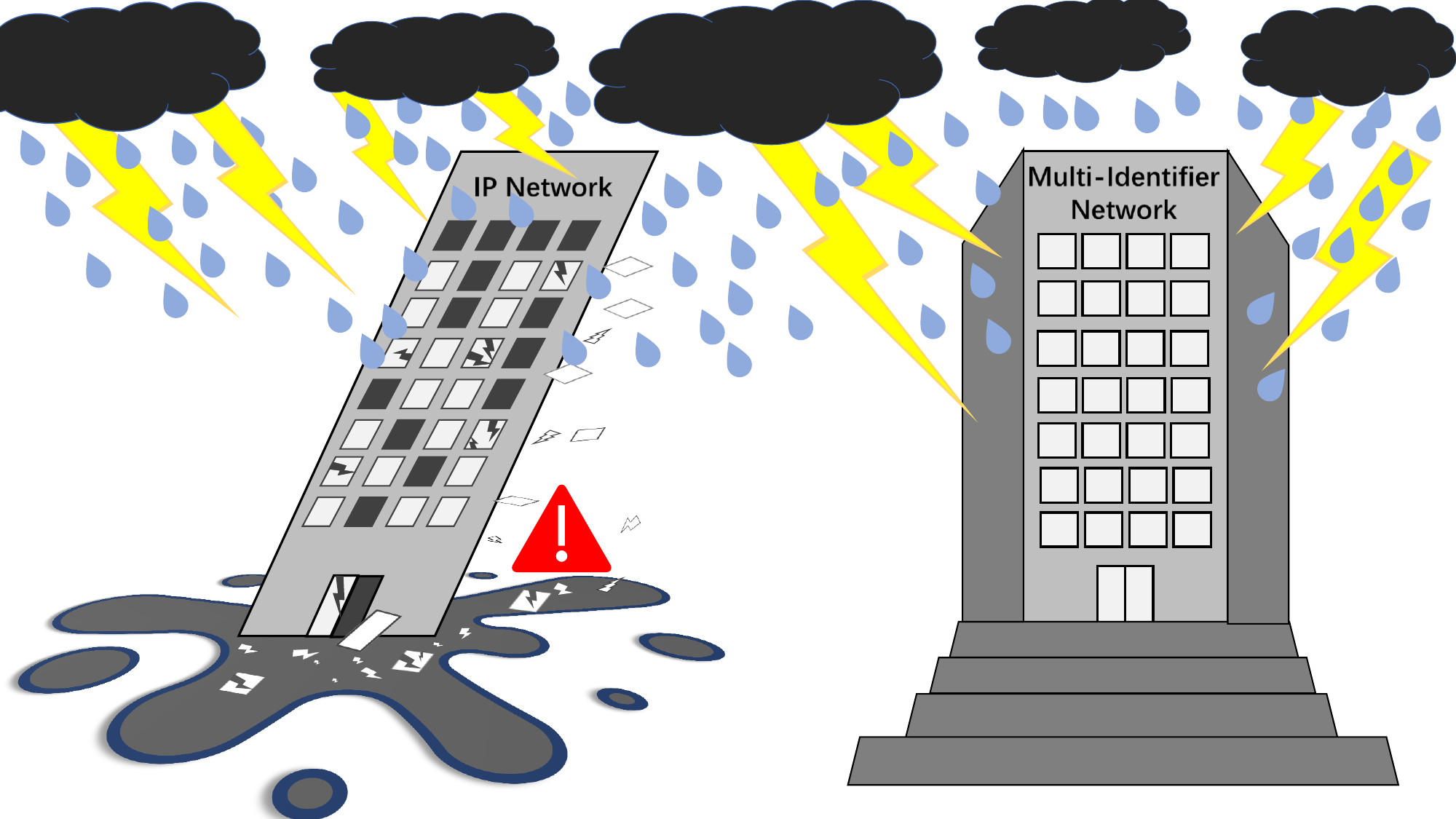}
    \caption{Security comparison between IP network and MIN network}
    \raggedright{\small Maintaining security in IP networks is analogous to building skyscrapers on unstable sand, susceptible to catastrophic failure under attack. Conversely, MIN's security resembles structures built on solid bedrock, resilient to extreme disruptions.} 
    
    \label{fig:sec}
\end{figure}

As shown in Figure~\ref{fig:sec}, maintaining security in IP networks is analogous to building skyscrapers on unstable sand, susceptible to catastrophic failure under attack. Conversely, MIN's security resembles structures built on solid bedrock, resilient to extreme disruptions. Future cyberspace architectures, such as MIN, offer deterministic security solutions that quantitatively and exponentially enhance defenses against network attacks:

\begin{enumerate}[\indent(a)]
    \item \textbf{Empirical Evidence}: Over the past three years, MIN has been tested as an independent “shooting range” in nearly 20 international security competitions, including the International Elite Challenge on Cyber Mimic Defense hosted by Zijinshan Laboratory. MIN remained consistently unbreached, demonstrating its superior resilience compared to IP systems.
    \item \textbf{Quantitative Analysis}: Using stochastic process models, such as the martingale framework, breaching a MIN-constructed private network is estimated to require approximately  $4.8 \times 10^{15}$  years under some normal scenario. For detailed proof, refer to Section 4.7.5 of the referenced monograph \cite{li2021co}.
\end{enumerate}

\subsection{Conjecture C: Deterministic security solutions may exist in future cyberspace}

\begin{figure}[htbp]
    \centering
    \includegraphics[width=0.7\linewidth]{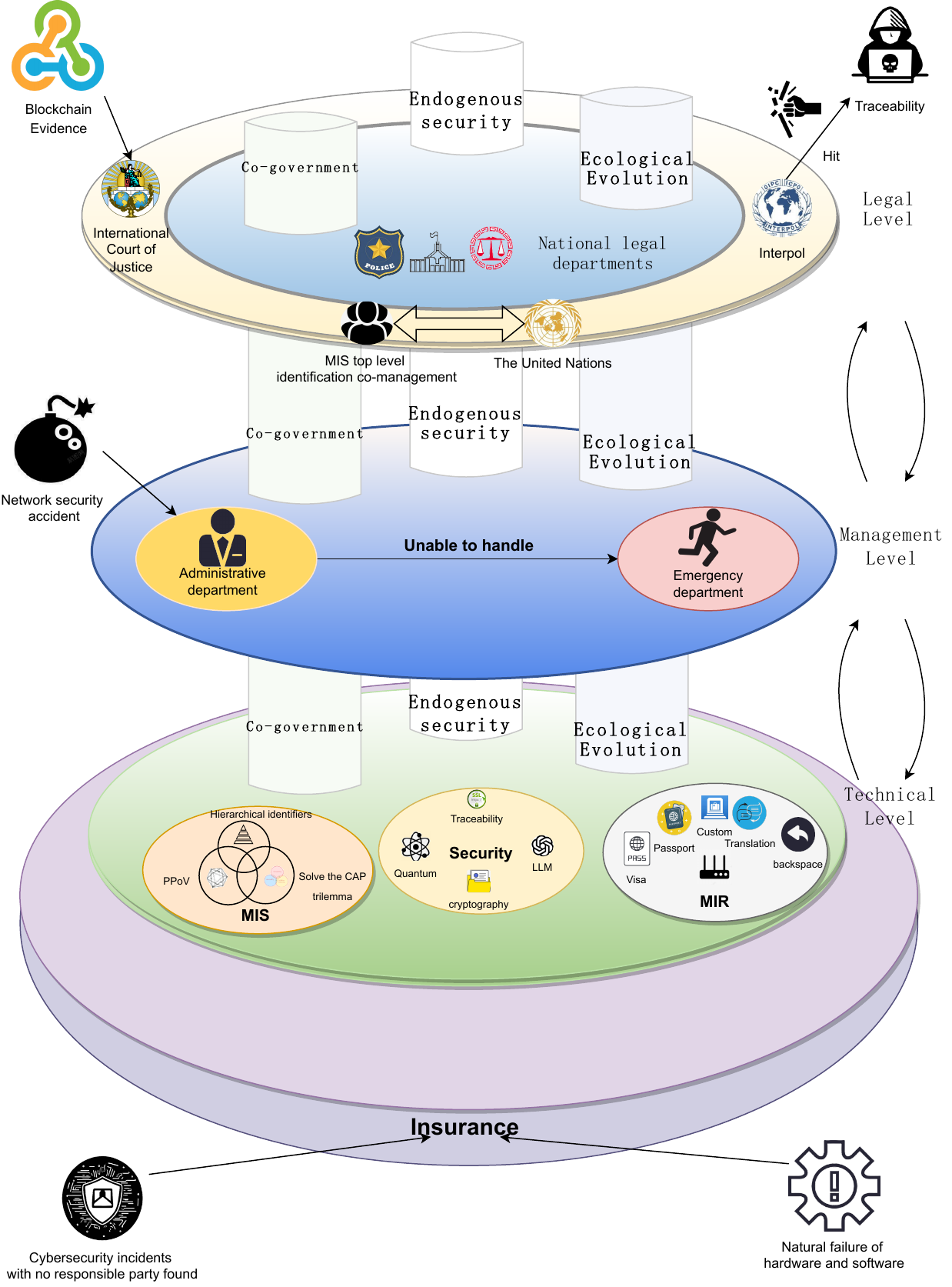}
    \caption{MIN from the technical level, management level and legal level of a complete set of solutions}
    
    \raggedright{\small MIN offers a multilayered solution that encompasses technical, management, legal, and insurance frameworks. At the technology level, the Multi-Identifier Router (MIR) and Multi-Identifier System (MIS) leverage advanced technologies to enhance network security, support global consensus, and disrupt monopolies, while the MIN architecture integrates multiple security features to transform cyberspace into a more equitable and secure environment. At the management level, when cyber incidents escalate beyond technical resolution, administrative and emergency departments intervene, potentially collaborating with law enforcement to trace malicious actors using blockchain records for legal action and compensation. At the legal level,  the establishment of a unified cyberspace court system utilizes blockchain technology to improve dispute resolution, enhance international cooperation, and prevent conflicts in the complex global digital landscape. Cyberspace insurance provides financial protection against data breaches and cyber incidents while leveraging the advanced security of the MIN network to offer competitive rates and promote ethical digital governance.
    
    } 
    \label{fig:threelayers}
    
\end{figure}

We conjecture that deterministic security solutions could be realized in future cyberspace, enabling a fundamental resolution to current vulnerabilities. Transitioning from the IP network architecture to high-security alternatives like MIN could serve as the foundation for such solutions. MIN offers a multilayered solution that encompasses technical, management, legal, and insurance frameworks, as illustrated in Figure~\ref{fig:threelayers}, promoting a rule of law, security, and peaceful cyberspace.

\subsubsection{Technical level}
At the technical level, the Multi-Identifier Router (MIR) replaces the traditional router, while the MIN architecture is poised to supersede the conventional IP network framework. The cyberspace customs mechanism embedded within MIR fortifies the security of national network borders, ensuring controlled and secure data flow between domains. Simultaneously, the identifier evolution and mutual translation mechanism supports the sustainable development of the network ecosystem by enabling seamless compatibility and adaptability across diverse architectures.

The Multi-Identifier System (MIS), underpinned by consortium blockchain technology, addresses the CAP trilemma inherent in distributed storage systems. By adopting the PPoV consensus mechanism, MIS supports large-scale global consensus, provides hierarchical identifier management on a worldwide scale, and effectively disrupts unilateral network monopolies, fostering a more equitable and decentralized cyberspace.

MIN's endogenous security design integrates multiple advanced features to ensure robust protection and adaptability. It safeguards content security through comprehensive network packet encryption, facilitates cybercrime prevention with detailed traceable records, and employs intelligent regulatory mechanisms powered by secure large-scale models. Furthermore, it incorporates quantum encryption, offering an unbreakable layer of security against unauthorized access. These innovations position MIN as a transformative architecture, driving the evolution of cyberspace toward enhanced rule of law, peace, and security.

\subsubsection{Management level}
When cyberspace encounters interference or attacks that cannot be effectively resolved through technical measures alone, administrative departments step in to provide manual intervention and support. Should the issue remain unresolved at this level, emergency departments are mobilized to address the network problems with urgency. In critical scenarios, management and emergency services may collaborate with national or international law enforcement agencies to ensure a comprehensive response.

Law enforcement agencies play a pivotal role by leveraging blockchain logs to identify and trace malicious users, enabling the swift cessation of cyberattacks and minimizing associated property losses. These blockchain records, with their immutable and tamper-proof characteristics, are subsequently submitted as evidence in court proceedings to facilitate litigation and secure reasonable compensation from those accountable for the damages. In cases of natural hardware or software failures, or rare incidents that cannot be traced, network insurance serves as a protective measure, providing compensation to safeguard the legitimate rights and interests of individuals and organizations in cyberspace.

\subsubsection{Legal level}
At the legal level, the complexities of modern global cyberspace necessitate the creation of a robust cyber dispute management mechanism. The traceability features of the MIN network empower international law enforcement agencies to efficiently track and neutralize malicious actors. To complement this capability, we propose the establishment of a unified cyberspace court system built on the Multi-Identifier System (MIS) and global blockchain logs. The immutable and tamper-proof nature of blockchain logs provides reliable evidence for legal arbitration, ensuring fairness and accountability in resolving disputes.

This proposed system addresses critical limitations inherent in the current Internet Protocol (IP) framework, particularly its challenges in evidence collection and its lack of capacity to effectively support transnational trials. By implementing a cyberspace adjudication system, countries can collaboratively resolve disputes, reduce the risk of conflict escalation, and prevent disagreements from escalating into severe international incidents or warfare.

Our approach envisions a globally recognized, enforceable legal framework, seamlessly integrated with the technical architecture of MIN. This framework encourages nations to resolve disputes through established legal channels rather than resorting to unilateral actions. We advocate for the United Nations to lead the development of an international cyberspace law grounded in legal principles and supported by blockchain-based evidence. Such an initiative would not only enhance global cooperation but also lay the foundation for lasting peace and stability in cyberspace.

\subsubsection{Insurance level}
Cyber insurance serves as a financial safeguard against the costs incurred from data breaches, cyberattacks, and other cyber-related incidents. It also provides coverage for natural hardware and software failures or other untraceable issues, such as bit dislocation, by facilitating claim settlements and ensuring property protection. The exceptional security of the MIN network significantly reduces the likelihood of internal security incidents, creating a promising opportunity to promote cyber insurance as a complementary service.

Insurance institutions can capitalize on the advanced security features of the MIN network to offer highly competitive insurance rates, incentivizing broader adoption of cyber insurance policies. In addition, these institutions can contribute to the sustainability of the network by paying reasonable fees to support the maintenance and enhancement of the MIS and related systems.

Promoting cyber insurance seamlessly aligns with the principles of ethical and responsible digital governance. By enforcing stringent security measures and adhering to robust regulatory standards, insurance companies play an active role in fostering a more secure, resilient, and reliable cyberspace, which benefits both individuals and organizations.

\section{Conclusion}
The evolution of IP networks, from their origins as a free and open system to their current state dominated by centralized monopolies, underscores the need for profound reflection on the future of network governance, security, and technological innovation. This paper has conducted an in-depth analysis of the history and current challenges of IP networks, systematically identifying three critical issues: centralized governance, fragile security mechanisms, and architectural rigidity that stifles innovation.

To address these challenges, we have introduced the Co-governance Multi-Identifier Network (CoG-MIN) architecture. MIN not only provides theoretical solutions to these pressing issues but has also demonstrated remarkable performance and immense practical potential. The Multi-Identifier System (MIS) leverages decentralized consortium blockchain technology to democratize network governance, enhance user security, and significantly improve traceability, thereby mitigating the risks of cyberattacks.

In addition, MIN ensures full compatibility with existing IP protocols while offering a flexible and adaptive ecosystem for future network evolution. By supporting diverse network architectures and identifiers, MIN lays a robust foundation for fostering the diversity and sustainable development of global networks. Its innovative design has attracted widespread attention and recognition from the international community.

The future of cyberspace requires concerted efforts and collaboration to shape it into a secure and inclusive environment. The MIN architecture provides a transformative framework for achieving a rule-of-law, peaceful, secure, and ecologically evolving network ecosystem. Two of the largest telecom operators in China have adopted MIN as their technology solution. As MIN progresses toward becoming a global standard for telecommunications carriers, it is poised to play a pivotal role in shaping the next generation of networked systems. We look forward to working with the global community to bring this vision to fruition and create a more secure and equitable cyberspace for all.

\bibliographystyle{unsrt}  
\bibliography{references}  
\end{document}